\documentstyle[preprint,aps]{revtex}
\begin {document}
\draft
\preprint{UCI-TR 94-10}
\title{Neutrino Fluxes and Resonance Physics with Neutrino Telescopes}
\setcounter{footnote}{0}
\author{Myron Bander\footnote{Electronic address:
mbander@funth.ps.uci.edu}\addtocounter{footnote}{3}%
}
\address{
Department of Physics, University of California, Irvine, California
92717, USA}
\author{H. \ R. \ Rubinstein\footnote{Electronic address: rub@vana.physto.se }}
\address{
Department of Theoretical Physics, University of Uppsala, Uppsala,
Sweden\\and\\Department of Theoretical Physics, Hebrew University,
Jerusalem, Israel
}
\date{March\ \ \ 1994}
\maketitle
\begin{abstract}
Expected atmospheric ${\bar \nu_e}$ fluxes will result in a
significant number of
resonantly produced low mass hadronic vector states, as the $\rho$, in
large volume
neutrino telescopes.  The existence of sources of higher energy neutrinos will
result in the production of higher mass states, as the $D_s^*$ and the $({\bar
t}b)_{J=1}$. We calculate the rates of production of these states and
discuss their
signals. Independent of theoretical flux determinations, the detection of these
states will be a tool for experimentally determining these fluxes.
\end{abstract}

Large volume high energy neutrino telescopes are coming into operation
with effective areas of the order of  $10^9$ cm${}^2$ and volumes greater than
$2\times 10^{13}$ cm${}^3$ \cite{amanda,dumand,baikal}.  Planned is the NESTOR
\cite{nestor} detector with a potential volume of  $10^{14}$ cm${}^3$.  In this
note we point out a method of determining or setting bounds on
${\bar\nu_e}$ fluxes
at various energies corresponding to the production of the standard vector
quark-antiquark resonances, namely $\rho$, $D_s^*$ and possibly $({\bar t}b)$;
rates for the production of other states is significantly lower
because of Cabibbo
or helicity suppression.  This complements the Glashow \cite{glashow}
mechanism for
resonant $W^{-}$ production.

A guaranteed source, with energies below $10^4$ GeV, are atmospheric
neutrinos
\cite{lipari}.  Neutrinos in this energy range will be capable of producing the
$\rho$ meson and the $D_s^*$. The extrapolated, to higher energies, atmospheric
neutrino fluxes give totally negligible rates for the production of higher mass
resonances.  However, interesting sources of neutrinos with energies
higher than
$10^4$ GeV have been postulated; Active Galactic Nuclei (AGN)
\cite{agn} are the
most promising source in that TeV gamma rays have been detected from
Markarian 421
\cite{markarian} and there is consensus that fluxes of all neutrino flavors of
comparable intensity exist \cite{protheroe}; as a large number of AGNs
is known to
exist one can estimate the expected diffuse neutrino flux \cite{stecker}.
Unexpected sources might be early universe relic neutrinos  \cite{berezinski}.

Some time back Glashow\cite{glashow} pointed out that the W boson can
be produced
in a resonant way in neutrino-electron scattering. We study the resonance
production
of hadronic states: $\rho$'s, $D_s^*$'s with a mass of 2.1 GeV and of the $J=1$
state of the $({\bar t}b)$
system.  For completeness we also present results for the helicity
suppressed $\pi$
production and the Cabibbo suppressed $K^*$ production.  For a vector
meson $R$ the
rate in a volume $V$ is \begin{equation}
 \mbox{\rm Rate}=\frac{48\pi^3\Gamma(R\rightarrow e\nu)}{M_Rm_e}
    \Phi\left (\frac{M_R^2}{2m_e}\right )N_eV\, ,\label{rate1}
\end{equation}
where $N_e$ is the elctron density (in water $N_e=3.4\times
10^{23}$/cm${}^3$) and $\Phi(E)$ is the flux, averaged over azimuthal
angles, at ${\bar\nu}_e $ energy E. In the above we assume that the
flux does not
change rapidly over the width of the resonance.

The partial width $\Gamma\left [(Q{\bar q})_{J=1}\rightarrow
e\nu\right ]$, for a
vector resonance made out of $Q$ and ${\bar q}$ quarks, is obtained
from the related
electromagnetic of the $Q{\bar Q}$ system. For the latter we use the empirical
relation \cite{ioffe}
\begin{equation}
\Gamma\left [(Q{\bar Q})_{J=1}\rightarrow e^-e^+\right ]=12e_Q^2\mbox{\rm
keV}\, ,
\label{emwidth}
\end{equation}
where $e_Q$ is the charge of the $Q$ quark. In terms of the
nonrelativistic quark
model this implies that the wave function at the origin is proportional to the
reduced mass. Using these facts we find
\begin{equation}
\Gamma\left [(Q{\bar q})_{J=1}\rightarrow e\nu\right ]=192M_Q^2M_q^2
\left (\frac{G_F}{4\pi\alpha}\right)^2\mbox{\rm keV}\, .\label{gamma1}
\end{equation}
For the $\rho$ meson conservation of isospin gives a firmer result, consistent
with
the one above,
\begin{equation}
\Gamma(\rho\rightarrow e\nu)=12M_{\rho}^4
\left (\frac{G_F}{4\pi\alpha}\right)^2\mbox{\rm keV}\, .\label{gammarho}
\end{equation}
Combining Eq. (\ref{rate1}) and Eq. (\ref{gamma1}) we find for a volume of
$2\times 10^{13}$ cm${}^3$ (the smallest of the proposed neutrino telescope
volumes)
\begin{equation}
\mbox{\rm Rate}=8\times 10^{11} \frac{M_Q^2M_q^2}{M_R}\Phi\left
(\frac{M_R^2}{2m_e}\right )/
\mbox{\rm year}\, ,\label{rate2}
\end{equation}
where $\Phi$ is in units of (cm${}^2$ s sr GeV)${}^{-1}$ and all
masses are in GeV.
For the $\rho$ meson the corresponding result is
\begin{equation}
\mbox{\rm Rate}=1.7\times 10^{10}\ \Phi(580\ \mbox{\rm GeV})/\mbox{\rm year}\,
{}.
\label{rhorate}
\end{equation}
The results, with calculated atmospheric neutrino fluxes (ATM)
\cite{lipari} and
theoretically estimated active galactic neutrino fluxes (AGN)
\cite{stecker,halzen}
are given in Table \ref{result1}. For convenience we present the flux
needed to
obtain 10 events/year. (For the calculations we use a $t$ quark mass
of 175 GeV.)

We also present, in Table \ref{result2}, results for the production
of two low mass
states that are either helicity suppressed $\pi$ or Cabibbo
suppressed $K^*$.
We note that at $E=6.4\times 10^6$ GeV, the $W^-$ production rate is 32/year.

The atmospheric neutrino flux calculations \cite{lipari} are
conservative and we
certainly expect that the neutrino telescopes will see several $\rho$
events per
year. The $\nu_{\mu}$ fluxes are calculated to be an order magnitude
larger. Should
there be any significant neutrino mixing \cite{imb} enhancing the $\nu_e$
flux, the rates due to atmospheric neutrinos would go up significantly. As
mentioned
earlier the rates presented in this work are for the smallest volume telescope
planned. Although the rate for production of $({\bar t}b)_{J=1}$ due to the
calculated AGN neutrino flux is well below the feasibility of any telescope,
there might be totally unexpected sources. The detection of hadronic resonances
will provide an experimental determination or limit on ${\bar\nu}_e$ fluxes. We
have
presented rates for events totally contained in the neutrino telescope volume.
These
events will be characterized by having no visible particle entering the volume
and
600 GeV or more of hadronic energy deposited locally in the detector in thin
hadronic
and/or electromagnetic shower of length 6 m to 10 m. \cite{detection}.

M.\ B.\ was supported in part by the National Science Foundation under
Grants PHY-9208386 and INT-9224138. H.\ R.\ was supported by the
Swedish Research Council and an EEC Science grant. We wish to thank
Dr. S.\ Barwick, Dr.\ S.\ Carius and Mr. Mats Thunman for valuable
discussions.

\nobreak

\begin{table}
\caption{Event Rate for Unsuppressed Vector Meson Production in $2\times
10^{13}$
cm$^3$.}
\begin{tabular}{cccccc}
State&Energy(GeV)&ATM Flux&AGN Flux&Events/year&Flux for 10/yr.\\
\tableline
$\rho$&$5.8\times 10^2$&$6\times 10^{-11}$&{}&1.2&$5\times 10^{-10}$\\
$D_s^*$&$4.4\times 10^3$&$10^{-13}$&{}&0.03&$3\times10^{-11}$\\
$({\bar t}b)_{J=1}$&$3.2\times 10^7$&{}&$5\times 10^{-21}$&$2\times
10^{-5}$&$2.5\times 10^{-15}$\\
\end{tabular}\label{result1}
\end{table}

\begin{table}
\caption{Event Rate for Suppressed Meson Production in $2\times 10^{13}$
cm$^3$.}
\begin{tabular}{ccccc}
State&Energy(GeV)&ATM Flux&Events/year&Flux for 10/yr.\\
\tableline
$\pi$&$19.4$&$10^{-5}$&0.03&$3\times 10^{-3}$\\
$K^*$&$7.8\times 10^2$&$6\times 10^{-11}$&0.06&$8\times10^{-9}$\\
\end{tabular}\label{result2}
\end{table}
\end{document}